\documentclass{sigchi}
\usepackage{balance}  
\usepackage{graphics} 
\usepackage[T1]{fontenc}
\usepackage{txfonts}
\usepackage{mathptmx}
\usepackage[pdftex]{hyperref}
\usepackage{color}
\usepackage{booktabs}
\usepackage{textcomp}
\usepackage{microtype} 
\usepackage{ccicons}  

\usepackage{todonotes}

\def\plaintitle{Stakeholder Involvement:\\
A Success Factor for Achieving Better UX Integration}

\def\emptyauthor{}
\def\plainkeywords{Authors' choice; of terms; separated; by
  semicolons; include commas, within terms only; required.}

\makeatletter
\def\url@leostyle{%
  \@ifundefined{selectfont}{
    \def\UrlFont{\sf}
  }{
    \def\UrlFont{\small\bf\ttfamily}
  }}
\makeatother
\def\pprw{8.5in}
\def\pprh{11in}

\definecolor{linkColor}{RGB}{6,125,233}
\hypersetup{%
  pdftitle={\plaintitle},
  pdfauthor={\emptyauthor},
  pdfkeywords={\plainkeywords},
  bookmarksnumbered,
  pdfstartview={FitH},
  colorlinks,
  citecolor=black,
  filecolor=black,
  linkcolor=black,
  urlcolor=linkColor,
  breaklinks=true,
}

\newcounter{rpracno}
 \DeclareRobustCommand{\rprac}[1]{%
 	\refstepcounter{rpracno}%
 	\textbf{\therpracno }\ \label{#1}}
 
 \newcounter{rpracnob}
 \DeclareRobustCommand{\rpracb}[1]{%
 	\refstepcounter{rpracnob}%
 	\textbf{\therpracnob}\ \label{#1}}

\usepackage{etoolbox}
\makeatletter
\def\@copyrightspace{\relax}
\makeatother

\newcommand{\libpath}{library.bib}

\begin{document}

\title{\plaintitle}

\numberofauthors{3}
{
\author{%
	\alignauthor{Pariya Kashfi\\
		\affaddr{Chalmers University of Technology}\\
		\affaddr{Gothenburg, Sweden}\\
		\email{pariya.kashfi@chalmers.se}}\\
	\alignauthor{Kati Kuusinen\\
		\affaddr{University of Central Lancashire}\\
		\affaddr{Preston, United Kingdom}\\
		\email{kkuusinen@uclan.ac.uk}}\\
 	\alignauthor{Robert Feldt\\
 		\affaddr{Chalmers University of Technology}\\
 		\affaddr{Gothenburg, Sweden}\\
 		\email{robert.feldt@chalmers.se}}\\
}

\maketitle

\begin{abstract}
 Stakeholder involvement is one of the major success factors in integrating user experience (UX) practices into software development processes and organizations. 
 It is also a necessity for agile software development.
 However, practitioners still have limited access to guidelines on successful involvement of UX stakeholders in agile settings. 
 Moreover, agile UX literature does not well address the specific characteristics of UX and it does not clearly differentiate between UX and usability work. 
 This paper presents two guidelines for supporting stakeholder involvement in both UX integration and the daily UX work.
 In particular, we focus on the special characteristics of UX: being dynamic, subjective, holistic, and context-dependent. 
 The guidelines clarify practical implications of these characteristics for practitioners. In addition, they can help researchers in addressing these characteristics better in agile UX research.
\end{abstract}

\section{Introduction}
The overall experience of a user with a software system is called User eXperience (UX).
The ISO definition of UX is as follows: a ``person's perceptions and responses resulting from the use and/or anticipated use of a product, system or service''~\cite{ISO9241}. 
UX is subjective, context-dependent, and dynamic, and holistic~\cite{Hassenzahl2010a}.

Compared to usability, which is often seen as a necessary precondition for good UX~\cite{Hassenzahl2010a,Lallemand2015b}, UX is more subjective and faster to change and thus more difficult to measure as such. 
Moreover, the perception of UX is generally different in academic and industrial contexts: whereas the former concentrates on hedonic aspects and emotions, the latter focuses more on functionality and usability issues~\cite{Vaananen2008a}.

UX work has its roots in User-Centered Design (UCD) which emphasizes stakeholder, in particular end user, involvement in development projects~\cite{ISO9241}.
Thus, involving stakeholders is inherent in UX work.
Agile processes share the people-centeredness and iterativity with UX work. 
However, although the idea of agile processes is to involve all stakeholder roles~\cite{Highsmith2001a} most of the popular agile methodologies, such as Scrum, recognize only the role of customers and not, for instance, end users. 
In particular, whereas UCD emphasizes early understanding of end users and the context, agile relies on little upfront design and instead chunks the design into short development iterations~\cite{Salah2014a}.
Moreover, agile methodologies are developer-centric and forget many other relevant roles such as UX experts. 
Thus, UCD and agile have different understanding of when and to what extent stakeholders should be involved in projects which creates challenges in UX work in agile settings~\cite{Kuusinen2012a,Salah2014a}.   

In this paper we approach the problem of stakeholder involvement from two perspectives. 
First, to address stakeholder involvement in UX work, companies require best practices for integrating UX principles and practices into their development processes.
Moreover, successful integration necessitates also in organizational matters, such as roles and responsibilities and organizational culture~\cite{Ferreira2011a,Rohn2012a}. 
Second, companies should concentrate on the interplay between UX experts, developers and other stakeholders during the actual design and development.
We present our early results on stakeholder involvement that are found through empirical studies in a number of software development companies (for more details, see~\cite{Kashfi2016b,Kashfi2016a}).

\vspace{-8pt}
\section{Background}
\label{sec:bg}

Current UX models  (e.g.~\cite{Hassenzahl2003a,Wright2010a}) differ in their view on how various underlying elements and processes contribute to forming the end user's overall experience with products and services.
One of the well-known UX models is developed by Hassenzahl~\cite{Hassenzahl2003a}.
It breaks UX down into \textit{pragmatic} and \textit{hedonic} attributes.
End user's perception of these attributes leads to a judgment about the product's appeal (e.g., ``It is good/bad''), emotional consequences (e.g., pleasure, satisfaction) and to behavioral consequences (e.g., increased time spent with the product). 
While pragmatic attributes concern achieving \textit{do-goals}, hedonic attributes concern satisfying \textit{be-goals}~\cite{Hassenzahl2003a}.
Do-goals are the concrete outcome that the end user wishes to achieve whereas be-goals rest in essential human needs.

Although practitioners cannot guarantee a specific experience, certain practices can increase the likelihood of delivering good UX~\cite{Hassenzahl2010a}. 
We refer to these practices as \textit{UX practices}. 
They ensure that be-goals, hedonic attributes of UX, and users' emotional reactions and consequences are taken into account in development of software systems~\cite{Hassenzahl2010a,Hassenzahl2001a,Law2014b}.
These practices include for instance the following: identifying relevant be-goals and creating design solutions that reflect those~\cite{Wright2010a}, understanding and identifying relevant UX measures that reflect users' perception on hedonic and pragmatic aspects of UX~\cite{Law2014b}, and refining abstract be-goals to more concrete do-goals and product quality characteristics~\cite{Hassenzahl2001a}.

However, simply applying UX practices in isolation is not enough.
Instead, these practices need to be integrated into development processes and considered throughout projects to make an impact.
In addition, UX principles (e.g. valuing essential human needs) should become an integral part of the organizational culture~\cite{Ferreira2011a,Kashfi2016a,Rohn2012a}.
We refer to the process of integrating UX principles and practices into development processes and organizations as \textit{UX integration}. 
UX integration is a socio-technical endeavor~\cite{Ferreira2011a}: it requires organizational (e.g. introducing new roles, adjusting business strategies) as well as technical changes (modifying development processes, introducing new tools and methods).
UX integration is known to be difficult and faced with various challenges~\cite{Ferreira2011a,Kashfi2016a}.
Still, practitioners have limited access to actionable guidelines on UX integration~\cite{Folstad2010a}.

UX integration can be considered a type of Software Process Improvement (SPI)~\cite{Kashfi2016b}.
The idea behind SPI is that development processes often need to be managed and improved in order for the outcome software to be of better quality~\cite{Niazi2005a}.	
Therefore, and SPI body of knowledge can inspire UX integration research and practice.
In particular, since the software development community already has access to ample SPI guidelines and best practices~\cite{Niazi2005a}.
Since SPI guidelines are generic and they do not directly address UX or other specific software quality characteristics~\cite{Kashfi2016b}, they need to be customized to better suit UX integration.

\section{Method}
\label{sec:method}
Our research goal was to create two guidelines to support practitioners in improving UX work in their companies.
The \textit{guideline on practicing UX} is a list of UX specific characteristics and their practical implications for the daily work of UX staff members, managers and non-UX staff members  including development and marketing.
This guideline is based on Hassenzahl's model that presents UX as a dynamic, subjective, holistic and context-dependent phenomenon~\cite{Hassenzahl2010a}, and the author's empirical studies on UX work in software companies~\cite{Kashfi2016a}.
The \textit{guideline on UX integration} is a list of specific practices that practitioners need to perform in order to better involve stakeholders and successfully integrate UX into their organizations.
This guideline is inspired by existing guidelines on how to improve success of SPI efforts in general~\cite{Niazi2005a}.
In addition, we discuss the implications of the characteristics of UX for stakeholder involvement and differentiate it from stakeholder involvement in SPI and in usability integration.

Both guidelines are syntheses of previous empirical research on UX integration by two of the authors, that is currently in submission/revision.
In our previous work~\cite{Kashfi2016a}, we identified challenges practitioners face in their work with UX, and reflected on how differences between UX and usability can intensify UX challenges.
To deepen our understanding of UX challenges, and also identify facilitators to UX integration, we then performed a longitudinal case study in a software development company (partially published in~\cite{Kashfi2016b}) and specifically focused on how the case company has moved from only developing user interfaces to also considering usability and more recently UX.
Here, we have applied our previous findings to more reflect on the issue of stakeholder involvement, and generate actionable guidelines for practitioners.

\section{Results}
\label{sec:result}

\subsection{A Guideline on Practicing UX}
This section presents our guideline on practicing UX.
This guideline includes a list of  UX characteristics and describes their implications on every day work of UX and non-UX staff members, also on management support.
It also includes the implications of these characteristics for software design and development, as summarized in Table~\ref{tbl:charac}.

\subsubsection{Dynamic Nature of UX and its Practical Implications}
UX is known to be dynamic, aka. temporal, emerging and changing over time~\cite{Hassenzahl2010a}. 
Therefore, in design and evaluation, UX staff members should know how to work with different episodes of UX~\cite{Law2014b} and companies require staff members with such knowledge.
Main episodes of UX are \textit{expected experience} (before usage), \textit{momentary experience} (during usage), \textit{remembered experience} (shortly after usage) and \textit{accumulated experience} (over longer period of use)~\cite{Hassenzahl2010a}. 

UX staff members need to reason which episodes are the most important for the software being developed.
For instance, for an e-marketing website first impression of users is more important  because the website's goal is to allure more visits. 
But in a work application accumulative experience is more important since users will interact with the application repeatedly over longer periods of time.
Based on relative importance of episodes of UX, UX staff members suggest design solutions and decide on the order of tasks users perform using the software.
These decisions may not be in agreement with how non-UX staff  members prefer to order the tasks based on other constraints (e.g. business goals, architecture).
For instance, UX staff  members may decide to remove the advertisements from the landing page of a website in order to improve the first impression of users.
This decision may be in conflict with the website's business goals.
Hence, management support is required to resolve potential conflicts between UX and non-UX staff members (e.g. requirement analysts, sales and marketing) considering the overlaps between their responsibilities.

In addition, UX staff  members also need to carefully design the user interaction since the order and timing of the tasks users perform impact various episodes of UX differently.
For instance, empirical studies show that remembered experience is mainly impacted by the last tasks a user performs~\cite{Hassenzahl2004b}. 
That is why user satisfaction questionnaires cannot truly measure the overall UX since they often mainly reflect the satisfaction of users from the last task performed~\cite{Hassenzahl2010a}.

Furthermore, to measure accumulative UX, UX staff members need to study longer periods of product and service usage after the system is released~\cite{Law2014b}.
It means that UX staff  members need support from management to set up strategies for this purpose, with close collaboration with sales and marketing units that also study longer usage of products and services.

In addition, dynamic nature of UX implies that expectations and initial judgments of users are formed before usage~\cite{Hassenzahl2010a}.
Therefore, the scene is to some extent set by the sales, marketing and business units when they advertise, negotiate and sell products to potential customers and users.
Hence, the role of these units is more important for UX compared to usability, since usability mainly concerns users interaction with the software.
This necessitates a close coordination and collaboration between UX staff  members and members of these  units.

\subsubsection{Subjective Nature of UX and its Practical Implications}
UX is subjective and heavily relies on human perception~\cite{Hassenzahl2010a,Wright2010a}.
They emphasize that objective qualities such as usability can be translated to subjective qualities, i.e. experienced or perceived qualities~\cite{Hassenzahl2010a}.
Usability, on the other hand, is a reflection of an objective approach to design which has its roots in cognitive psychology~\cite{Hassenzahl2010a}. 

For improving UX, enhancing objective qualities is not the only option or even the right one; rather, the right combination of various objective qualities can lead to a desired experience~\cite{Hassenzahl2010a}.
For instance, if a software system takes time to load, designers can improve the experience of users by introducing some tasks that can occupy users during the loading - it is shown that occupied time is perceived shorter than unoccupied time~\cite{Jones1996a}.
This is an example of a concept known as `experience pattern' or `transformation rule'~\cite{Hassenzahl2010a}.

By understanding such transformation rules (from the field of psychology), UX staff  members can create a suitable combination of objective qualities to increase the likelihood of the intended experience~\cite{Hassenzahl2010a}, balancing also with cost and feasibility.
For instance, shortening the loading time might be impossible due to high cost or technical limitations. Instead, its negative impact can be reduced by understanding this specific experience pattern.
Moreover, addressing the subjective perception of users, i.e. experiences, requires subjective metrics such as user surveys~\cite{Law2014b}.
 
Subjectivity of UX can lead to more power struggle, disagreements and conflicts among UX and non-UX staff  members.
First, everyone experiences different products and services every day.
Therefore, even non-UX staff  members can easily have opinions about what experiences a product should deliver, and how they should be delivered, i.e. through which design solutions.
Second, by understanding experience patterns, UX staff  members may suggest a specific combination of objective quality characteristics that is not necessarily in agreement with other stakeholders suggestions: e.g. a product owner may suggest a different set of quality characteristics to protect the customer's business goals. 
Hence, power struggles and disagreements may rise between UX and non-UX staff  members~\cite{Kashfi2016a,Kuusinen2012a}.

Admittedly, power struggle and disagreements are not unique to UX, but are often more difficult to resolve in this case~\cite{Kashfi2016a}.
In case of other quality characteristics, at least in theory, we can resolve these disagreements using objective evidence that shows why an alternative is better than another: i.e. through measurements. 
But measuring UX is more difficult than other quality characteristics and even impossible in earlier phases~\cite{Law2014b,Vermeeren2010a}.
Hence, management has an important role in resolving potential power struggles, disagreement and conflicts.
Although management may not have knowledge and expertise to decide between design alternatives, it can give UX staff  members enough authority to make the final decisions based on input from non-UX staff  members.
Management should also be committed and willing to support a process that has no clear or certain outcome or in business terms `return on investment'. 
A designer can never guarantee a type of experience because it relies on perception, but she can increase the likelihood of delivering it through careful design~\cite{Hassenzahl2010a}.

\subsubsection{Holistic Nature of UX and its Practical Implications}

UX work, in contrast to usability work, takes a holistic approach to design through focusing on creating specific experiences that support users' be-goals (i.e. psychological needs) as well as do goals~\cite{Hassenzahl2010a}. 
Therefore, UX staff  members need to have the knowledge and skills to understand the relevant be-goals and prioritize them, and to design for them accordingly which requires sufficient knowledge in human psychology.
In addition, be-goals need to be refined into concrete do-goals~\cite{Hassenzahl2001a}. 
Do-goals may also be identified by non-UX staff  members, e.g. requirement analysts.
These do-goals need to be combined and prioritized which requires constant communication and close collaboration between UX and non-UX staff  members to negotiate the do-goals and their relative importance.
Holistic nature of UX also implies more power struggle and disagreements between UX and non-UX staff  members similarly to what subjective and dynamic nature of UX does.

\subsubsection{Context-dependent Nature of UX and its Practical Implications}

UX research uses the term situated or context-dependent to emphasise that any experience is unique, unrepeatable, and  situated~\cite{Wright2010a}. 
Despite their uniqueness, experiences can be categorised because their essence is the same, i.e. they connect to essential human needs or be-goals~\cite{Hassenzahl2010a}.
Implications of being context-dependent largely overlaps the implications we discussed for holistic nature of UX.
For instance, UX staff  members need to understand these categories of experience, i.e. be-goals, and have the skills to design for particular groups of experiences.
Similarly, a close collaboration between UX and non-UX staff  members is needed, and a strategy to settle disagreements and overcome power struggles.
In addition, being context-dependent implies that field studies are more suitable for measuring and evaluating UX than lab studies~\cite{Law2014b}.
Field studies are known to result in more realistic data on experience, but they require more resources, hence requiring more management support.

\begin{table*}[h]
	\centering
	\scriptsize
	\begin{tabular}{|p{0.45\columnwidth}|p{0.5\columnwidth}|p{0.45\columnwidth}|p{0.45\columnwidth}|}
		\hline 
		\textbf{Dynamic} & \textbf{Subjective}  & \textbf{Holistic} & \textbf{Context-dependent} \\ \hline
		
		identify \& prioritize episodes of UX 
		
		{\tiny {\tiny {\tiny }}}	&
		combine \& decide on the order \& timing of do-goals based on experience patterns 
		&
		identify \& prioritize relevant be-goals
		&
		identify \& prioritize relevant be-goals
		
		\\ \hline 
		
		combine \& decide on order \& timing of do-goals based on episodes of UX 
		&
		evaluate based on user perception 
		&
		trade-off between be-goals \& do-goals from stakeholders other than users
		&
		refine be-goals to do-goals  
		\\ \hline
		
		trade-off between different combinations of do-goals with stakeholders other than users
		& 
		in evaluation pay attention to timing \& order to tasks users perform
		&  
		refine be-goals to do-goals
		&
		evaluate the intended types of experiences in the field  \\  \hline
		
		evaluate different episodes of UX
		
		&  make informed decisions based on evaluation of objective \& perceived qualities
		&
		evaluate whether the be-goals are satisfied
		&
		make informed decisions based on evaluations in labs and in the field  \\ 
		\hline 
	\end{tabular} 
	\caption{Implications of various characteristics of UX on requirements and design and evaluation activities in  software development.}
	\label{tbl:charac}
\end{table*}

\subsection{A Guideline on UX Integration}
Here we present our guideline on UX integration that consists of a list of practices that should be performed to increase the likelihood of successful UX integration.
Because of the scope of this paper, we have only included those practices that specifically relate to stakeholder involvement: staff (both UX and non-UX staff), and management.
Our guideline is inspired by existing guidelines on SPI, mainly the work of Niazi et al.~\cite{Niazi2005a} that shows how practitioners should address the known critical success factors and barriers to SPI in general.
Here, we have updated and, when applicable, adjusted the practices suggested for SPI success, to better suit UX integration.

\textbf{UX Integration Practices to Ensure Staff Involvement}

	\textbf{P}\rprac{}
	\textit{Promote the benefits of improving UX of products and services among the staff before starting UX integration.}
	The group who plans, initiates and executes UX integration, hereafter,\textit{ UX integration action group}, should inform the staff about why UX is required in addition to other quality characteristics (e.g., usability, security).

	\textbf{P}\rprac{} 
	\textit{Disseminate awareness among the staff concerning the importance and benefits UX integration.}
	In particular, the UX integration action group should inform the staff about why it is not sufficient to simply perform UX practices as add-on to current development processes.
	In addition, the staff should be informed about the differences between UX integration and usability integration, and that integrating usability practices is not sufficient to achieve better UX in product and services.
	
	\textbf{P}\rprac{} 
	\textit{Identify and ensure to involve and assign UX-related responsibilities to those staff members whose day-to-day work can directly or indirectly impact UX of the products or services.}
	For instance, product owners should assign part of their time for discussing feature designs with the UX staff  members, architects (one example of non-UX staff  members) should give input to and receive input from UX staff  members on how the intended UX design relates to the architectural choices.
	
	\textbf{P}\rprac{}   
	\textit{Inform the staff members about their or other staff members' roles and responsibilities in relation to UX practices, and UX integration practices.}
	For instance, developers should be informed that they should negotiate any changes to the user interface with the UX staff  members.
	They should also be informed that, because of UX integration, new roles (e.g. UX owner) are introduced with certain set of responsibilities and authorities.

	\textbf{P}\rprac{}  
	 \textit{Establish a mechanism to monitor the progress of staff in relation to UX integration}, specially, a mechanism to monitor how non-UX staff  members facilitate or prohibit UX practices and UX integration practices.
	 This is especially important to help identifying and addressing barriers to UX integration in the company.
    
	\textbf{P}\rprac{} 
	\textit{Ensure that UX integration reviews are regularly performed within the organization by the UX integration action group.}
	In each review meeting the participants should reflect on previous or ongoing UX integration practices, and discuss possible future adjustments.
	In these meetings, special attention should be paid to the following:
	(i) communication and collaboration between UX and non-UX staff  members,
	(ii) UX artifacts being generated and used,
	(iii) overlaps between roles and responsibilities of UX and non-UX staff  members.
    
	\textbf{P}\rprac{} 
	\textit{Involve all the key stakeholders in UX integration reviews.}
    Representatives of both UX and non-UX staff  members should participate to the review meetings.  
    It is in particular important to involve those staff members who might be less positive towards UX integration.
    Involving them can help better understanding their concerns and plan to address them.

	\textbf{P}\rprac{}
	\textit{Regularly organize events for increasing UX awareness within the organization. }
    For instance, UX integration action group
    can organize monthly internal workshops, or maintain Wiki pages to disseminate knowledge on UX, UX integration and their corresponding practices.
		
	\textbf{P}\rprac{}  
	\textit{Establish a UX integration action group with internal or external experienced practitioners.}
    Such practitioners should not only have knowledge on UX but also leadership, negotiation and communication skills.
    In particular, they should be able to negotiate with the non-UX staff members whose roles and responsibilities overlap with those of UX staff  members.
    
	\textbf{P}\rprac{}    
	\textit{Assign responsibilities to a number of staff members with the right authorities and skills to provide technical support to UX integration and UX practices.}
    For instance, UX integration action group requires a communication channel across organizational units that can be set up with the help of the staff members with required knowledge and authority.
    For instance, in an agile setting, a board on JIRA (an agile project management tool\footnote{\url{https://atlassian.com}}) can be assigned to UX integration issues with the help of product owners and scrum masters.
	
	\textbf{P}\rprac{}   
    \textit{Establish a mechanism to collect and analyze feedback from the UX integration action group and to reflect on the main lessons learned.}
    For instance, monthly retrospective meetings can be used  across the organization to reflect on experiences and lessons learned.
	
	\textbf{P}\rprac{}   
	\textit{Establish a process to distribute the lessons learned to the relevant staff  members.}
    For instance, reports from retrospective meetings should be sent to not only the participants but also all the UX staff  members across the organization to give them a better picture of the overall status of UX integration.

	\textbf{P}\rprac{}  
	 \textit{To perform UX integration and UX practices, involve those staff members who  have indicated interest and commitment to UX work.}
	 Involving motivated and interested staff members can increase the likelihood of UX integration success.
    
	\textbf{P}\rprac{}   
	\textit{Establish conflict resolution plans for potential conflicts between UX and non-XU staff  members}, in particular, concerning the conflicts that may arise when addressing be-goals and do-goals, or hedonic and pragmatic aspects of UX.
    This is especially important since UX is subjective and more difficult to measure than other quality characteristics.
    
	\textbf{P}\rprac{}    
	\textit{Allocate the necessary time to the staff members who perform UX integration or UX practices.}
	Ensure that both UX and non-UX staff members are happy with the allocated time, and none of them experiences time pressure.
	For instance, attending UX integration retrospective meetings should be a part of the staff's working hours, or specific time should be assigned to UX and non-UX staff members to negotiate the identified be-goals and do-goals resulted from them with requirements from other sources. 

	\textbf{P}\rprac{}   
	\textit{Ensure that UX integration practices do not get in the way of day-to-day work of practitioners.}
	This can be achieved through explicitly assigning time to UX integration practices.
	
	\textbf{P}\rprac{}  
	\textit{Involve representatives of all the key stakeholders in UX integration practices.}
	For instance, project management, sales and marketing should have representatives in planning UX integration.

	\textbf{UX Integration Practices to Ensure Management Support}

	\textbf{P}\rpracb{}  
	\textit{Promote benefits of UX among the management, at all levels of the organization.}
	 In particular, promote why moving beyond usability is beneficial for the company.

	\textbf{P}\rpracb{} 
	\textit{Ensure that management provides strong leadership and support for UX integration and UX practices.}
	
	\textbf{P}\rpracb{}  
	\textit{Ensure that management is committed to provide all the required resources for UX integration and UX practices.}
	In particular, managers should provide competences, time and budget to support hedonic aspects of UX.
	
	\textbf{P}\rpracb{}  
	\textit{Ensure that management establishes conflict resolution plans for potential conflicts arising among UX and non-UX staff members.}
	Most importantly because UX is more difficult to measure than other quality characteristics, and it is therefore more difficult to decide among design alternatives to address it.
	These resolution plans should in particular address potential power struggle and disagreements arising between UX staff members and developers and members of business units.
	
	\textbf{P}\rpracb{}  
	\textit{Ensure that management establishes UX practices as an integral part of the development process, and UX principles as part of the organizational culture.}
	In particular there is a need to establish practices for:
	(i) Requirement analysis: how to identify, prioritize and refine relevant be-goals
	(ii) Design: how to translate be-goals into design solutions
	(ii) Testing: how to evaluate hedonic aspects of UX, and overall UX of products and services in different phases
	
	\textbf{P}\rpracb{} 
	\textit{Ensure that management is willing to participate in UX integration assessment meetings and improvement workshops.}
	
	\textbf{P}\rpracb{}  
	\textit{Ensure that management is committed to provide training needed for UX integration.}
	In particular, the staff need to be aware of 
	(i) how UX relates to other quality characteristics, especially usability,
	(ii) why UX integration is important and goes beyond usability integration,
	(iii) why both hedonic and pragmatic aspects of UX are important.
	
	\textbf{P}\rpracb{} 
	\textit{Ensure to develop a process to review each critical success factor and critical barrier of UX integration.}
	
\section{Discussion}
\label{sec:disc}
Designing and developing for UX is a multidimensional and multidisciplinary activity which has not yet gained an established position in software industry.
Both approaches for UX integration and the daily cooperation between UX and non-UX staff members require further investigations. 
Even the UX requirements, design and evaluation methods are still developing, and introduced methods are not widely accepted in the industry.
Examples are emocards~\cite{Desmet2001a} for gathering users' momentary emotions about the interaction and UX curve~\cite{Kujala2011a} for long-term UX evaluation.
Although approaches such as service design~\cite{Goldstein2002a} have become more popular, usability and purpose of use still often dominate over designing for an experience~\cite{Kuusinen2012a,Vaananen2008a}. 
For instance, in Kuusinen et al's study~\cite{Kuusinen2012a} in a large software organization, ease of use and efficiency were the most often reported sources of good UX.

Current literature includes a number of guidelines to better integrate UX into software  organizations (e.g.~\cite{Salah2013a,Plonka2014a}).
But these guidelines often do not clearly separate UX and usability, therefore do not reflect on the implications of the differences between these two concepts. 
The particularities of UX- subjective, dynamic, holistic, and context dependent- require wider understanding of the end user compared to usability. 
We hope that the guidelines we have provided can address these shortcomings in the current literature on UX integration.

\section{Conclusion}
This paper presented two guidelines that can support improving UX work in software companies.
The guideline on practicing UX focuses on the daily work of UX staff  members and how  different characteristics of UX can impact this work.
In particular, it focuses on the collaboration between UX staff members and other internal stakeholders such as developers, marketing and management.
The guideline on UX integration focuses on introducing and adopting UX principles and practices into software development processes and organizations.
It consists of a list of supporting practices for involving both UX and non-UX staff members in UX integration and a second list for ensuring management support. 
In the future, we aim to evaluate our guidelines based on empirical data.
We also aim to extend the proposed practices by specific recommendations.
One example would be recommending a specific resolution plan that better suits common conflicts between UX and non-UX staff members.

\balance{}

\bibliographystyle{SIGCHI-Reference-Format}
\bibliography{\libpath}
\end{document}